\documentclass[preprint,showpacs,preprintnumbers,amsmath,amssymb,nofootinbib]{revtex4-1}

\usepackage{graphicx}
\usepackage{dcolumn}
\usepackage{bm}

\newcommand{\la}{\langle}
\newcommand{\ra}{\rangle}

\newcommand{\beq}{\begin{equation}}
\newcommand{\eeq}{\end{equation}}
\newcommand{\beqy}{\begin{eqnarray}}
\newcommand{\eeqy}{\end{eqnarray}}
\newcommand{\beqyn}{\begin{eqnarray*}}
\newcommand{\eeqyn}{\end{eqnarray*}}
\newcommand{\nl}{\newline}
\newcommand{\nn}{\nonumber}

\newcommand{\half}{\frac{1}{2}}

\newcommand{\bc}{\begin{center}}
\newcommand{\ec}{\end{center}}
\newcommand{\bmin}{\begin{minipage}}
\newcommand{\emin}{\end{minipage}}

\begin{document}

\title{A critical assessment of the angular momentum sum rules}

\author{Elliot Leader}
 \email{e.leader@imperial.ac.uk}
\affiliation{Blackett laboratory \\Imperial College London \\ Prince Consort Road\\ London SW7 2AZ, UK}

\date{\today}

\begin{abstract}
There are now five angular momentum  relations or sum rules in the literature: the Jaffe, Manohar relation for a longitudinally polarized nucleon, and the Bakker, Leader, Trueman result for the case of transverse polarization; the Ji relation for longitudinal polarization, and the Leader result for transverse polarization, both involving generalized parton distributions; and a new sum rule due to Ji, Xiong and Yuan dealing with the transverse component of the Pauli-Lubanski vector. I discuss these various relations and examine their precise interpretation in the light of the so-called ``angular momentum controversy". In particular, I show that the claim of Ji, Xiong, Yuan that their Pauli-Lubanski relation is frame or energy independent is incorrect and that they have missed an energy dependent term in their expression.
\end{abstract}
\pacs{11.15.-q, 12.20.-m, 12.38.Aw, 12.38.Bx, 12.38.-t, 14.20.Dh}

\maketitle

\section{\label{sec1}The Jaffe-Manohar and Bakker-Leader-Trueman  angular momentum relations}
For a \emph{longitudinally} polarized nucleon moving in the $Z$ direction the relation (or sum rule) between the angular momentum of the nucleon and the contributions to this from its constituent quarks and gluons, due to Jaffe and Manohar (JM) \cite{Jaffe:1989jz},
\begin{equation}
\half = \half \Delta \Sigma + \Delta G +
  L(q)  +  L(G)
\label{long}
\end{equation}
has been known for a long time. However, in the last few years, as a consequence of the controversy concerning the definition of quark and gluon angular momentum (for access to the literature see \cite{Leader:2011za}), it has been understood that care must be exercised in specifying which version of the angular momentum operators one is using. There are two principal versions: \nl

1) The \emph{canonical} angular momentum operators $\bm{J}^{can}$, which emerge directly from Noether's theorem, contain four terms: the quark spin $\bm{S}^{can}(q)$ and orbital angular momentum $\bm{L}^{can}(q)$ and the gluon spin $\bm{S}^{can}(G)$ and orbital angular momentum $\bm{L}^{can}(G)$. However, only one of these terms,  $\bm{S}^{can}(q)$, is gauge invariant.\nl

 2) The \emph{Belinfante} version of the angular momentum operators $\bm{J}^{bel}$ has only three terms, since the gluon angular momentum is not split into a spin and orbital part: the quark spin $\bm{S}^{bel}(q)=\bm{S}^{can}(q)$, the quark orbital angular momentum $\bm{L}^{bel}(q)$, and the gluon total angular momentum $\bm{J}^{bel}(G)$. The advantage of the Belinfante version is that all three terms are separately gauge invariant. The disadvantage is the absence of  a gluon spin term. \nl

Now the JM sum rule contains the term $\Delta G$, which is the first moment of the polarized gluon density $\Delta G(x)$. The latter is defined as the nucleon matrix element of a certain gauge invariant operator, which is \emph{not} the same as the gluon spin operator. However, it coincides with the canonical gluon spin operator \emph{in  the light-cone gauge} $A^+=0$. Thus the angular momentum appearing in the Jaffe-Manohar sum rule is actually
\beq \label{JMJ} \bm{J}^{JM} \equiv \bm{J}^{can}|_{\textrm{gauge} A^+=0}. \eeq
This should not be considered a restriction, because the parton model is really a ``picture" of QCD in the gauge $A^+=0$. So the first stage to stating the JM relation correctly, for a nucleon moving along $OZ$, is:
\begin{equation} \label{JMSR}
\half = \half \Delta \Sigma + \Delta G +
  L^{JM}_z(q)  +  L^{JM}_z(G)  \eeq
 where  $ L^{JM}_z(q,\, G)$ is the \emph{expectation value}, for a \emph{longitudinally} polarized nucleon:
 \beq \label{JML}  L^{JM}_z(q,\, G) \equiv \langle \langle \, L^{can}_z(q,\,G)\, |_{\textrm{gauge} A^+=0}\, \, \rangle  \rangle_L \eeq
 and $q$ means a sum over all quarks and antiquarks. However, there is  another subtlety.  The term $ \Delta \Sigma$ has to be extracted from a measurement, in deep inelastic scattering, of the polarized structure function $g_1(x, Q^2)$. But the relation between $g_1(x, Q^2)$ and $\Delta \Sigma$ is factorization scheme dependent, because what actually appears in $g_1(x, Q^2)$
 is the flavour-singlet axial charge $a_0$ (or $g^{(0)}_A$ ) and its connection with $\Delta \Sigma$ and $\Delta G$ is scheme dependent as a consequence of the axial anomaly (for a detailed discussion see \cite{Bass:1992ti} and \cite{Leader:1998nh}). This implies that each term in Eq.~(\ref{JMSR}) must be scheme dependent and should have a scheme label, including the orbital angular momentum, an issue which, to the best of my knowledge has not been studied.\nl

 On the basis of the analysis in  \cite{Jaffe:1989jz}, it was believed for many years that no such relation was possible for a \emph{transversely} polarized nucleon. However, in 2004, Bakker, Leader and Trueman (BLT) \cite{Bakker:2004ib} showed there was a subtle error in
\cite{Jaffe:1989jz}, which did not affect the longitudinal case, but rendered the analysis of the transverse case incorrect.  Subsequently, using a very careful wave packet analysis BLT, using the canonical version of the angular momentum, derived a sum rule for the transverse case,
\begin{equation}\label{eq.T}
 \half = \half\, \sum_{flavours}\, \int dx \, [ \Delta _T q (x)+ \Delta _T \bar{q} (x)]
+ \sum_{q,\, \bar{q}, \, G }\langle \langle \, L^{can}_T \, \rangle \rangle. \eeq
Two things should be noted in regard to this relation:\nl

1) In deriving this result the term $[\Delta _T q (x)+ \Delta _T \bar{q} (x)]$ arose from the Fock expansion of the nucleon state. Now usually the transversity densities $\Delta _T q (x)$  originate from the matrix element of $\bar{\psi}\,\gamma^+ \,\gamma^1 \, \gamma_5 \, \psi $, which is a chiral-odd operator. Since the angular momentum operator is chirally-even it has been suggested that Eq.~(\ref{eq.T}) cannot be correct. However, the Fock expansion expression which emerges from the nucleon matrix element of the operator $\bar{\psi}\,\gamma^+ \,\gamma^1 \, \gamma_5 \, \psi $ is, stricly, the \emph{difference} $[\Delta _T q (x)- \Delta _T \bar{q} (x)]$. Thus the terms in Eq.~(\ref{eq.T}) involving the transversity density are \emph{not}  linked to this operator, and thus the criticism based on chirality seems to be invalid. I have rechecked the analysis leading to Eq.~(\ref{eq.T}) and  believe that it is indeed correct.\nl

2) Eq.~(\ref{eq.T}) is valid in an ``infinite momentum frame" i.e. assumes that the nucleon momentum $P\rightarrow \infty $. \nl
For finite mementum there is a correction term related to the gluon helicity polarization inside a transversely polarized nucleon. Including this, for a nucleon with momentum $P$ in the $Z$ direction, Eq.~(\ref{eq.T}) should be replaced by
\beqy \label{Tcor}  \half &=& \half\, \sum_{flavours}\, \int dx \, [ \Delta _T q (x)+ \Delta _T \bar{q} (x)] + \sum_{q,\, \bar{q}, \, G }\langle \langle \, L^{can}_T \, \rangle \rangle \nn \\
&+& \frac{\pi}{M}\int dx k_T dk_T \left[\sqrt{x^2P^2 + k_T^2} -xP \right] \, g^G_{1T}(x, k_T^2)
 \eeqy
 where the gluon helicity polarization in a nucleon polarized, for example,  in the $X$  direction is given by
 \beqy \label{Ghpol} \Delta G_{h/s_x}(x, \bm{k}_T) & \equiv & G_{1/s_x}(x, \bm{k}_T) - G_{-1/s_x}(x, \bm{k}_T)  \nn \\
 &=& \frac{k_x}{M}g^G_{1T}(x, k_T^2) \eeqy
 where $g^G_{1T}(x, k_T^2)$ is the gluon analogue of the function $g_{1T}(x, k_T^2)$ introduced in \cite{Mulders:1995dh}. \nl
 Note that both the JM and BLT relations have the disadvantage that they only yield information on the \emph{sum} of quark and gluon orbital angular momentum.\nl
I have stressed that  in  the above relations the angular momentum corresponds to the \emph{canonical definition}, where the angular momentum operators are the generators of rotations. Different, and more  interesting relations hold for the Belinfante version of the angular momentum operators, to which I shall now turn.
\section{ \label{sec2} The Ji and Leader angular momentum relations}
Some time ago Ji \cite{Ji:1997pf, Ji:1996ek, Ji:1996nm}, using the \emph{Belinfante version} of the angular momentum operator, derived an expression for the expectation value of the \emph{longitudinal} component of the quark angular momentum   in a \emph{longitudinally} polarized nucleon
\beq \label{JiL} J^{Ji}_z(q) \equiv \langle \langle \,\, J_z^{bel}(q)\, \,\rangle  \rangle_L  \eeq
in terms of the Generalized Parton Densities (GPDs) $H$ and $E$, namely (often this relation is misleadingly written without the label specifying the longitudinal component)
\beq \label{Ji}  J_z^{Ji}(q)  =\frac{1}{2 }\, \left[ \int_{-1}^{1} dx x E_q(x,0,0) +  \int_{-1}^{1} dx x H_q(x,0,0) \right] \eeq
 where here, as earlier,  and in the following, ``$q$" signifies the contribution of a quark \emph{plus} antiquark of a given flavor. An analogous result holds for gluons.\nl

More recently \cite{Leader:2011cr} I derived  similar relations for the instant-form \emph{transverse} components of the quark and gluon angular momentum in a \emph{transversely} polarized nucleon
\beq \label{JT}\langle  \langle \, J_T^{bel}(q) \, \rangle \rangle_T =\frac{1}{2M }\, \left[ P_0\,\int_{-1}^{1} dx x E_q(x,0,0) + M \, \int_{-1}^{1} dx x H_q(x,0,0) \right]   \eeq
and
\beq \label{JTG} \langle \langle \, J_T^{bel}(G) \, \rangle \rangle_T=\frac{1}{2M }\, \left[ P_0\,\int_{0}^{1} dx x E_G(x,0,0) + M \, \int_{0}^{1} dx x H_G(x,0,0) \right]   \eeq
where $P_0$ is the energy of the nucleon (note that for gluons the integrals run from $0$ to $1$). It was explained that the, at first sight, unintuitive energy dependence is, in fact, quite natural from a classical point of view (see Section 2.6 of \cite{Landau:1951}).
Of course the energy dependence cancels out when adding the quark and gluon contributions, so that
\beq \sum_{flavors}\langle \langle \, J_T^{bel}(q) \, \rangle \rangle_T + \langle\langle \, J_T^{bel}(G) \, \rangle\rangle_T = \frac{1}{2}. \eeq
This follows because the coefficient of $P_0$ vanishes i.e. because
\beq \label{EqG} \sum_{flavors} \int_{-1}^{1} dx x E_q(x,0,0) + \int_{0}^{1} dx x E_G(x,0,0) =0 \eeq
which is a  fundamental sum rule with wide ramifications and can be shown to correspond to the vanishing of the nucleon's anomalous gravitomagnetic moment \cite{Brodsky:2000ii, Teryaev:XXX}.
Now Ji, Xiong and Yuan (JXY) \cite{Ji:2012vj} have criticized the above on the grounds that it is, what they call ``frame dependent", meaning that the split between the quark and the gluon contributions depends upon the energy of the nucleon. As mentioned above, this energy dependence seems to be quite natural. Nonetheless JXY have claimed  to derive a relation which is frame independent, based upon the use of the Pauli-Lubanski vector. I shall argue in the next Section that their claim is not justified.
\section{ \label{sec3} The Ji-Xiong-Yuan transverse polarization relation}
Whether one is studying the Pauli-Lubanski vector,
\beq W_\mu = \half \epsilon_{\mu\rho\sigma\nu}\,J^{\rho\sigma}\,P^\nu \eeq
where $P^\nu $  and  $ J^{\rho\sigma}$ are the momentum and angular momentum operators and the convention is
$\epsilon_{0123}= 1 $,
or  the angular momentum itself, one requires an expression for the expectation value of the angular momentum operator. Note that actually of  the six independent  antisymmetric ``angular momentum" operators, only three, namely $J^{ij}, i,j=1,2,3$ are genuine angular momentum operators, while the other three $J^{0i}$ are actually boost operators. The six operators are given in terms of the angular momentum density tensor $M^{\mu\rho\sigma}(x)$ by
\beq \label{J} J^{\rho\sigma}= \int \bm{d^3x}\, M^{0\rho\sigma}(t, \bm{x}). \eeq
It is important to note the following conventions which I shall use throughout this section:\nl
1)  As is done by JXY, I shall be using the \emph{Belinfante} version of the momentum and angular momentum  operators, so this  will not be indicated  explicitly \nl
2) an operator \emph{without} a quark or gluon label ($q$ or $G$)  always refers to the total system i.e. the nucleon itself \nl
3) I am using the \emph{instant} form of dynamics where the system evolves as a function of time $t$. The results do not change if one uses light-front dynamics.
 \subsection{Expressions for the matrix elements of the angular momentum tensor}
The Belinfante angular momentum density tensor is given in terms of the energy momentum density tensor by
\beq \label{EMT} M^{\mu\rho\sigma}(x)= x^\rho \,T^{\mu\sigma} - x^\sigma \,T^{\mu\rho} \eeq
and the problem is that substituting Eq.~(\ref{EMT}) into Eq.~(\ref{J}) yields an integral which  is highly ambiguous:
\beqy  \int \bm{d^3 x} \langle P \, | x^i T^{0j} (x) - x^j T^{0i} (x)
| P  \rangle
   \nonumber \\
   &  & \hspace{-2cm} = \int \bm{d^3 x} \,
   x^i \,\langle P \,  | e^{i P \cdot x} T^{0j} (0) e^{-i
P \cdot x} | P \, \rangle
   - (i\leftrightarrow j)
   \nonumber \\
   &  & \hspace{-2cm} = \int \bm{d^3 x} \, x^i  \,\langle P \, | T^{0j}(0) | P \, \rangle
    - (i\leftrightarrow j).
\label{ambiguous}
\eeqy
Since the matrix element in Eq.~(\ref{ambiguous}) is independent of $x$ the integral  is totally ambiguous, being
either infinite or, by symmetry, zero.
There are two ways to deal with this: using  a careful wave-packet analysis as was done by BLT, or using a trick due to JM, which is the approach followed by JXY, but which, as will be explained,  is potentially dangerous. \nl
Thus JXY begin their analysis in the same way as Jaffe-Manohar \cite{Jaffe:1989jz} by considering the quantity
\beq  {\mathcal M}^{\mu\nu\lambda} (P,k,P',\mathcal{S}) \equiv \int d^4 x e^{i k\cdot x}
   \langle P', \mathcal{S} | M^{\mu\nu\lambda} (x) | P, \, \mathcal{S} \rangle \label{JM} \end{equation}
   which is designed to regulate the ambiguity in Eq.~(\ref{ambiguous}).  The state is specified by its 4-momentum $P$ and the covariant spin vector $\mathcal{S}$, normalized to $\mathcal{S}^2= - M^2$.\nl
   One may wonder why a 4-dimensional Fourier Transform is introduced in dealing with a 3-dimensional integral. The reason is that ${\mathcal M}^{\mu\nu\lambda} (P,k,P',\mathcal{S})$  \emph{appears} to transform as a covariant Lorentz tensor, but that is an illusion, because the non-forward matrix element of a tensor operator is not a tensor. This misunderstanding is partly responsible for the error in \cite{Jaffe:1989jz} mentioned above, but is avoided by JXY.
Note also that for
the validity of the later manipulations  in JXY $P'$ should be treated as an independent variable and should not be put equal to $P +k$. Note too, that strictly speaking, the covariant 4-vector $\mathcal{S}$ in the final state cannot be the same as in the initial state, since for a physical nucleon one must have $P \cdot \mathcal{S}= P' \cdot \mathcal{S} =0 $. The correct way to handle Eq.~(\ref{JM}) is to specify the same \emph{rest frame vector} $\bm{s}$ in both initial and final states. This should be understood in the following.\nl
Eventually  the limits $k^\mu \rightarrow 0$ and $P'\rightarrow P$ are taken, and this will yield an expression for the matrix elements of $J^{\rho\sigma}$ up to a trivial factor.   \nl

Now the genuine angular momentum of the system is a conserved operator i.e. commutes with the hamiltonian and does not contain any explicit factors of $t$, so that
\beq J^{ij}(t, \bm{x})= J^{ij}(0 , \bm{x})\qquad  \textrm{is independent of}\quad t. \eeq
Although the boost operators $J^{0j}$ do not commute with the hamiltonian, and contain an explicit factor of $t$,  they too are independent of time:
\beq \label{boost} J^{0j}(t, \bm{x})= J^{0j}(0 , \bm{x}). \eeq
Then
\beqy \lim_{\bm{k}\rightarrow 0} \lim_{P'\rightarrow P}\, \mathcal{M}^{0\nu\lambda} (P,k,P',\mathcal{S})&= & \int \, dt \, e^{ik_0t} \,\int \bm{d^3 x}
   \langle P,\, \mathcal{S} | M^{0\nu\lambda} (x) | P, \, \mathcal{S} \rangle \nn \\
   &=& 2\pi \,\delta (k_0) \,\langle P,\, \mathcal{S} | J^{\nu\lambda} | P, \, \mathcal{S} \rangle  \eeqy
   so that finally one can access the matrix elements of $J^{\nu\lambda}$ via
\beq \label{limJij}   \lim_{k^\mu \rightarrow 0} \lim_{P'\rightarrow P}\, \mathcal{M}^{0\nu\lambda} (P,k,P',\mathcal{S})= 2\pi \,\delta (0) \,\langle P, \, \mathcal{S} | J^{\nu\lambda} | P, \, \mathcal{S} \rangle \eeq
   and the innocuous factor $2\pi \,\delta (0)$ cancels out later in the calculation. \nl
In summary the JM trick is very useful provided one is careful not to give $\mathcal{M}^{0\nu\lambda} (P,k,P',\mathcal{S})$ incorrect transformation properties.\nl
As shown by the wave-packet analysis of BLT, $\mathcal{M}^{0\nu\lambda} (P,k,P',\mathcal{S})$  contains two terms, one referring to the internal angular momentum of the nucleon, the other to the angular momentum of the packet about the origin of coordinates. Only the first term is of interest and after some manipulation yields, for the spatial indices $j,k$
\beq \label{Jfinal} \frac{\langle P, \mathcal{S}\, |\, J^{jk}\, | P, \, \mathcal{S} \, \rangle}{i (2\pi)^3 \delta^3(0 )} = \left[\frac{\partial}{\partial  \Delta_j}  \langle \, P+\Delta/2,\, \mathcal{S}\, |\, T^{0k } \, |\, P-\Delta/2, \, \mathcal{S}\, \rangle - (j \leftrightarrow k) \, \right]_{\Delta^\mu \rightarrow 0}. \eeq
For the indices $0,k$ one finds
\beq \label{Jfinal1} \frac{\langle P, \mathcal{S}\, |\, J^{0k}\, | P, \, \mathcal{S} \, \rangle}{i (2\pi)^3 \delta^3(0 )} = -\left[\frac{\partial}{\partial  \Delta_k}  \langle \, P+\Delta/2,\, \mathcal{S}\, |\, T^{00 } \, |\, P-\Delta/2, \, \mathcal{S}\, \rangle  \right]_{\Delta^\mu \rightarrow 0}. \eeq
Note that care must be exercised in taking the derivatives since, for a physical nucleon, $P\cdot \Delta =0$ so that $\Delta_0= \frac{\bm{P}\cdot \bm{\Delta}}{P_0}$, and $\Delta_0$ is thus not an independent variable.\nl
It is clear from Eq.~(\ref{Jfinal}) that all we require is an expansion of \nl $\langle \, P+\Delta/2,\, \mathcal{S}\, |\, T^{\mu\nu}\, |\, P-\Delta/2, \, \mathcal{S}\, \rangle $ to first order in $\Delta$, but this is a little tricky and was given incorrectly in \cite{Jaffe:1989jz}. The  result used by JXY agrees with the expression derived  for the first time by BLT \cite{ Bakker:2004ib}, namely, in the notation of JXY\footnote{The connection with  the notation in \cite{Leader:2011cr} is : $A=\mathbb{D}/2, B=\mathbb{S} -\mathbb{D}/2$},
\beqy \label{Tmunu}\langle \, P+\Delta/2,\, \mathcal{S}\, |\, T^{\mu\nu}\, |\, P-\Delta/2, \, \mathcal{S}\, \rangle &= &2(AP^\mu P^\nu +  M^2\bar{C} g^{\mu\nu})\nn \\
   && \hspace{-8cm}i \frac{\Delta_\rho}{M}\left\{ \frac{A + B}{2}[P^\mu \epsilon^{\rho\nu\alpha\beta} + P^\nu \epsilon^{\rho\mu\alpha\beta}] +
 \frac{AP^\mu P^\nu + M^2 \bar{C} g^{\mu\nu}}{P_0 + M}\epsilon^{0\rho\alpha\beta} \right\} \frac{\mathcal{S}_\alpha}{M} P_\beta \eeqy
where we have written $A, B, \bar{C}$ for $A(\Delta^2=0), B(\Delta^2=0), \bar{C}(\Delta^2=0)$ respectively, and have included the term  $ \bar{C} g^{\mu\nu}$ which is only present if $T^{\mu\nu}$ is a \emph{non-conserved} operator i.e. when using Eq.~(\ref{Tmunu}) for the individual quark and gluon contributions.

Now taking
 $P^\mu = (P_0,0,0,P)$ and comparing the expression for the matrix elements of $T^{++}_{q,G}$ obtained from Eq.~(\ref{Tmunu}) with the expression given by Diehl for the GPDs \cite{Diehl:2003ny} one obtains the connection between $A$ and $B$ and the GPDs (details can be found in \cite{Leader:2011cr}):
  \beq \label{Hsum} A_q = \, \int_{-1}^{1} dx x H_q(x,0,0) = \,\left[ \frac{P_q}{P} \right] \eeq
  where $P_q =  \textrm{ momentum carried by a quark plus antiquark of a given flavour}$, and
  \beq \label{Esum}B_q =  \,\int_{-1}^{1} dx x E_q(x,0,0). \eeq
  Similar relations hold for the gluon contribution.
Using these in Eq.~(\ref{expJ}) and choosing $\mathcal{S}$ to correspond to a longitudinally or transversely polarized nucleon  one obtains the relations quoted in Eqs.~(\ref{Ji}) and (\ref{JT}, \ref{JTG}).
\subsection{Approach via the Pauli-Lubanski vector}
As mentioned, JXY are critical of the energy dependence of the relations Eq.~(\ref{JT}, \ref{JTG}) and suggest that this can be avoided by considering the transverse component of the Pauli-Lubanski vector rather than just the angular momentum itself. But, as I shall show, there are  problems with this approach and I believe the JXY results are incorrect.
Now both the Belinfante momentum and angular momentum operators are \emph{additive} in the sense that
  \beq \label{add} P^\mu =\sum_{flavours} P^\mu_q + P^\mu_G  \qquad \textrm{and} \qquad J^{\mu\nu} =\sum_{flavours} J^{\mu\nu} _q + J^{\mu\nu}_G .\eeq
Clearly, then, the Pauli-Lubanski vector is not additive
  \beqy \label{notadd}  W_\mu &=& \half \epsilon_{\mu\rho\sigma\nu}\,J^{\rho\sigma}\,P^\nu \\ & =&  \half \epsilon_{\mu\rho\sigma\nu}\,\left[\sum_{flavours} J^{\rho\sigma} _q + J^{\rho\sigma}_G \right] \, \left[ \sum_{flavours} P^\nu_q + P^\nu_G \right] \nn \\
 &  & \neq  \sum_{flavours} W_\mu (\textrm{quark}) + W_\mu(\textrm{gluon})\eeqy
In order to work with  additive quantities, JXY \emph{define} operators which they call $W_\mu(\textrm{quark})$ and $W_\mu(\textrm{gluon})$, but which are not really the quark and gluon Pauli-Lubanski vectors since they contain the \emph{total} momentum operators:
 \beq \label{PLq} W_\mu^{JXY}(\textrm{quark})=-\half \epsilon_{\mu\rho\sigma\nu}\,J^{\rho\sigma}_q\,P^\nu \eeq
 and
 \beq \label{PLG} W_\mu^{JXY}(\textrm{gluon})=-\half \epsilon_{\mu\rho\sigma\nu}\,J^{\rho\sigma}_G\,P^\nu \eeq
 so that
 \beq \label{Wconts} W_\mu = \sum_{flavours}W_\mu^{JXY}(\textrm{quark}) + W_\mu^{JXY}(\textrm{gluon}). \eeq
The question then is what is the physical interpretation of these operators? It is known on general grounds that the expectation value of the \emph{total} Pauli-Lubanski vector has a very simple and direct meaning in terms of the covariant spin vector specifying the state (see Section 3.4 of \cite{Leader:2001gr}) namely, for a  spin 1/2 particle\footnote{An analogous relation holds for arbitrary spin: see \cite{Leader:2001gr}}
 \beq \label{ExPL} \langle\langle P, \mathcal{S}\, |\, W^\mu\, | P, \, \mathcal{S} \, \rangle\rangle = \half \mathcal{S}^\mu \eeq
 where
 \beq \label{Sdef} \mathcal{S}^\mu = \left(\bm{P}\cdot \bm{s}, \, M \bm{s} + \frac{\bm{P}\cdot \bm{s}}{P_0 + M}\, \bm{P} \right) \eeq
 and $\bm{s}$ is the \textrm{rest frame} spin vector. \nl
 Clearly, then, for example,
 \beq W^\mu_ {JXY}(\textrm{quark}) \neq \half \mathcal{S}^\mu(\textrm{quark}) \eeq
 but
 \beqy W^\mu _{JXY}(\textrm{quark}) & \equiv & \half \mathcal{S}^\mu(\textrm{nucleon})_q \nn \\
 & \equiv & \textrm{contribution of $q$ to $\half \mathcal{S}^\mu(\textrm{nucleon})$ } .\eeqy
Thus
 \beq \mathcal{S}^\mu(\textrm{nucleon})= \sum_{flavours}\mathcal{S}^\mu(\textrm{nucleon})_q + \mathcal{S}^\mu(\textrm{nucleon})_G. \eeq
Let us now calculate the expectation value of the transverse component of the \emph{total} Pauli-Lubanski vector
\beq W_x = W^1 = \,[P_0\,J^{23} + P \, J^{02}]. \eeq
Using Eq.~(\ref{Tmunu}) in Eqs.~(\ref{Jfinal}) and (\ref{Jfinal1}) and dividing by $2P_0$ we obtain for the expectation values of  the $J^{\mu\nu}$.

\beq \label{PM02}
 \la \la  P,  \mathcal{S}  | J^{02}  |  P, \mathcal{S}  \ra \ra  = \frac{1}{P}\left[ \frac{M -P^0}{2M}\,A -\frac{P^2}{2M^2}\,B\right]\mathcal{S}_x.
\eeq
and
\beq \label{P0M23}
 \la \la  P, \mathcal{S} | J^{23}  |  P, \mathcal{S} \ra \ra = \frac{1}{P_0}\left[\frac{P_0}{2M}\,A+ \frac{(P^0)^2}{2M^2}\,B\right]\mathcal{ S}_x,
\eeq
Note that since I am dealing with the total angular momenta at this point, there is no contribution from the $\bar{C}$ terms,  and one finds,
\beq \label{Wexp} \langle\langle P, \mathcal{S}\, |\, W_x | P, \, \mathcal{S} \, \rangle\rangle = \half (A+B)\,\mathcal{S}_x. \eeq
We can check whether this is consistent with Eq.~(\ref{ExPL}). Using Eq.~(\ref{Hsum})
\beq \label{A} A = \sum_{flavours}A_q + A_G = \, \left[\frac{\sum_{flavours}P_q + P_G}{P}\right] = 1 \eeq
while, from Eqs.~(\ref{Esum}) and (\ref{EqG})
\beq \label{B} B = \sum_{flavours}B_q + B_G = 0. \eeq
Thus Eq.~(\ref{Wexp}) becomes
\beq \langle\langle P, \mathcal{S}\, |\, W_x | P, \, \mathcal{S} \, \rangle\rangle = \half \, \mathcal{S}_x \eeq
as expected.
Repeating this calculation for $\langle\langle P, \mathcal{S}\, |\,\,  W^{JXY}_x(\textrm{quark})\, \,| P, \, \mathcal{S} \, \rangle\rangle $ I obtain
\beq \label{WJXYq} \langle\langle P, \mathcal{S}\, |\,\,  W^{JXY}_x(\textrm{quark})\, \, | P, \, \mathcal{S} \, \rangle\rangle = \left[ \half(A_q + B_q) + \frac{P_0 - M}{2P_0}\, \bar{C}_q \right]\, \mathcal{S}_x. \eeq
Note that the term $\bar{C}_q$ appears here because $J^{\mu\nu}_q$ is not a conserved operator. A similar relation holds for the gluon contribution.
It follows from Eqs.~(\ref{Hsum}, \ref{Esum}) and the analogous relations for gluons that
\beq \label{SumSq} \mathcal{S}^\mu(\textrm{nucleon})_q =  \frac{1}{2}\,\left[ \int_{-1}^{1} dx x H_q(x,0,0) +  \int_{-1}^{1} dx x E_q(x,0,0)\right] + \frac{P_0 - M}{2P_0}\, \bar{C}_q \eeq
and
\beq \label{SumSG} \mathcal{S}^\mu(\textrm{nucleon})_G =  \frac{1}{2}\,\left[ \int_{0}^{1} dx x H_G(x,0,0) +  \int_{0}^{1} dx x E_G(x,0,0)\right] + \frac{P_0 - M}{2P_0}\, \bar{C}_G  \eeq
which disagrees with the relations given in JXY, which do not contain the energy dependent terms involving $\bar{C}_q$ and $\bar{C}_G$.\nl
JXY comment that because
\beq \label{additive} J^{\mu\nu} = \sum_{flavours}J^{\mu\nu}_q + J^{\mu\nu}_G \eeq
is a conserved operator and therefore that
\beq \label{sumC} \sum_{flavours}\bar{C}_q + \bar{C}_G =0, \eeq
 it is legitimate to ignore the $\bar{C}$ terms in Eqs.~(\ref{SumSq}, \ref{SumSG}). I do not see how this observation can justify leaving out the $\bar{C}$ terms in Eqs.~(\ref{SumSq}, \ref{SumSG}).
 Since the individual quark and gluon angular momenta are definitely not conserved, it is certain that neither $\bar{C}_q$ nor $\bar{C}_G$ can be zero. Indeed I have shown that $\bar{C}_q$ is related to the GPDs defined in \cite{meissner:2009ww} by\footnote{A proof is given in section VI H of \cite{LeaderXXX}.}
 \beq \label{Cbar} \bar{C}_q=  \int_{-1}^1 xdx(H_{3,\,q} - \half H_q), \eeq
 and the relation Eq.~(\ref{SumSq}) becomes
 \beqy \label{SumSqfinal}
\mathcal{S}^q_x(\textrm{nucleon}) &= & \frac{1}{2}\left[ \frac{P^0 + M}{2P^0}\int_{-1}^{1} dx \,  x\, H_q(x,0,0) +  \int_{-1}^{1} dx \,  x\, E_q(x,0,0)\right. \nn \\
 & & \left. + \frac{P^0-M}{P^0}\int_{-1}^{1} dx\,  x\, H_3^q(x,0,0)\right].
\eeqy
 Thus the claim of JXY that they have produced a frame independent relation is not valid. In fact one sees from this and Eq.~(\ref{SumSG}) that the JXY result is correct only at $P_0=M$, i.e. \emph{in the rest frame} !
\section{Summary}
It is now widely recognized that the definition of quark and gluon angular momentum is far from unique and several variants have been proposed, of which, arguably, the \emph{canonical} and \emph{Belinfante} versions are the most important. Sum rules or relations based on these are equally interesting and  explore different aspects of the internal spin structure of the nucleon. I believe that the relations based on the \emph{canonical} angular momentum, namely the Jaffe, Manohar result for longitudinal polarization and the Bakker, Leader, Trueman relation for transverse polarization are correct, despite an apparent contradiction with chirality in the BLT case. Concerning the relations based on the \emph{Belinfante} version, the Ji result for longitudinal polarization and the Leader relation for  transverse polarization are certainly correct, even though in the latter case the proportion of angular momentum contributed by quarks and gluons is energy dependent. I have, however, shown that the new relation of Ji, Xiong and Yuan, utilizing the transverse component of the Pauli-Lubanski vector, which attempts to circumvent such an energy dependence, is, unfortunately, incorrect. The corrected results \emph{are} frame or energy dependent and it turns out that the relations given by Ji, Xiong and Yuan  actually only hold in the rest frame of the nucleon.
\section{Acknowledgements}
I am grateful to Ben Bakker, Marcus Diehl, Xiangdong Ji and Feng Yuan for comments and to Feng Yuan for sending me a draft copy of the JXY paper. I thank C\'{e}dric Lorc\'{e} for certain clarifications derived from the draft copy of his recent paper \cite{Lorce:XXX}.
\section{Note added}
 Hatta, Tanaka and Yoshida \cite{Hatta:XXX} have also noted that the JXY  relation  is not frame independent.

\bibliography{Elliot_General}

\end{document}